\documentclass[%
 reprint,
superscriptaddress,
 amsmath,amssymb,
 aps,
]{revtex4-2}

\usepackage[utf8]{inputenc}
\usepackage{graphicx}
\usepackage{color}
\usepackage{physics}
\usepackage{braket}
\usepackage{comment}
\usepackage{mathrsfs}
\usepackage{enumerate}
\usepackage[version=4]{mhchem}
\usepackage[breaklinks=true]{hyperref}
\newcommand{\mr}[1]{\mathrm{#1}}

\newcommand{\mcl}[1]{\mathcal{#1}}
\newcommand{\bbC}{\mathbb{C}}
\newcommand{\bbR}{\mathbb{R}}

\newcommand{\bbN}{\mathbb{N}}
\newcommand{\sign}{\mathrm{sign}}

\newcommand{\poly}[1]{\mathrm{poly} \left( #1 \right)}
\date{\today}
\usepackage{amsthm}
\theoremstyle{definition}

\newtheorem*{theorem*}{Theorem}
\newtheorem*{proposition2*}{Proposition}

\begin{document}
\title{Recursive Quantum Eigenvalue/Singular-Value Transformation: \\ Analytic Construction of Matrix Sign Function by Newton Iteration}

\author{Kaoru Mizuta}
\email{mizuta@qi.t.u-tokyo.ac.jp}
\affiliation{RIKEN Center for Quantum Computing (RQC), Hirosawa 2-1, Wako, Saitama 351-0198, Japan}
\affiliation{Department of Applied Physics, The University of Tokyo, Hongo 7-3-1, Bunkyo, Tokyo 113-8656, Japan}

\author{Keisuke Fujii}
\affiliation{Graduate School of Engineering Science, Osaka University, 1-3 Machikaneyama, Toyonaka, Osaka 560-8531, Japan.}
\affiliation{Center for Quantum Information and Quantum Biology, Osaka University, Japan.}
\affiliation{RIKEN Center for Quantum Computing (RQC), Hirosawa 2-1, Wako, Saitama 351-0198, Japan}
\affiliation{Fujitsu Quantum Computing Joint Research Division at QIQB,
Osaka University, 1-2 Machikaneyama, Toyonaka 560-0043, Japan}

\begin{abstract}
Quantum eigenvalue transformation (QET) and its generalization, quantum singular value transformation (QSVT), are versatile quantum algorithms that allow us to apply broad matrix functions to quantum states, which cover many of significant quantum algorithms such as Hamiltonian simulation.
However, finding a parameter set which realizes preferable matrix functions in these techniques is difficult for large-scale quantum systems: there is no analytical result other than trivial cases as far as we know and we often suffer also from numerical instability.
In this Letter, we propose recursive QET or QSVT (r-QET or r-QSVT), in which we can execute complicated matrix functions by recursively organizing block-encoding by low-degree QET or QSVT.
Owing to the simplicity of recursive relations, it works only with a few parameters with exactly determining the parameters, while its iteration results in complicated matrix functions.
In particular, by exploiting the recursive relation of Newton iteration, we construct the matrix sign function, which can be applied for eigenstate filtering for example, in a tractable way.
We show that an analytically-obtained parameter set composed of only $8$ different values is sufficient for executing QET of the matrix sign function with an arbitrarily small error $\varepsilon$.
Our protocol will serve as an alternative protocol for constructing QET or QSVT for some useful matrix functions without numerical instability.
\end{abstract}

\maketitle

\textit{Introduction.---}
Quantum Eigenvalue Transformation (QET) is a versatile quantum algorithm which enables to apply various matrix functions \cite{Gilyen2019-qsvt}.
For a hermitian matrix of interest $A$, QET executes parallel processing of its eigenvalues and thereby allows us to apply broad matrix polynomial functions $\sum_n c_n A^n$ to arbitrary quantum states.
With its generalization to general matrices, called quantum singular value transformation (QSVT), it covers various today's important quantum algorithms such as Hamiltonian simulation \cite{Low2017-qsp,Low2019-qubitization} and search algorithms \cite{Grover1997-prl}, by properly constructing polynomial approximations.
Not only it provides unified understanding of quantum algorithms \cite{Martyn2021-grand-unif}, but also it can serve more efficient alternative algorithms for various purposes.

QET can be executed by repetition of parameterized unitary gates on ancilla qubits and unitary gates embedding the target matrix, called block-encoding.
While tunability of the parameters ensures realization of broad functions by quantum signal processing (QSP) \cite{Low2017-qsp}, we must accurately determine a proper parameter set for a desired function.
Although finding the parameters for degree-$q$ polynomials within an error $\varepsilon$ can be executed by $\poly{q,\log (1/\varepsilon)}$-time classical computation, its numerical instability has become one of the central problems for accurate implementation of QET/QSVT.
In fact, several numerical algorithms trying to solve this instability, such as parameter determination by optimization, have recently been developed \cite{Haah2019-numerical_QSP,Rui2020_numerical_QSP,Dong2021-symmetric_QSP,Wang2022-symmetric_QSP}.
By contrast, there are only a few results on analytical parameter determination.
As far as we know, they are limited to trivial cases for Chebyshev polynomials, which are useful for Grover's search algorithm \cite{Grover1997-prl,Hoyer2000-grover,Long2001-grover}.

In this Letter, we propose recursive QET/QSVT (r-QET/r-QSVT) that can potentially determine all the parameters in an analytical or numerically-much-cheaper way.
In these protocols, we recursively organize block-encoding by low-degree QET/QSVT so that it can reproduce recursive relations of matrix functions, and then obtain complicated matrix functions by iteration.
For instance, we can exploit Newton iteration for matrix functions as the recursive relation \cite{Higham2008-matrix}.
Then, with a sufficient number of iterations for its convergence, r-QET/r-QSVT enables us to organize nontrivial matrix functions only with a smaller number of parameters that can be easily determined.
As a prominent consequence, we obtain a tractable implementation of matrix sign functions with arbitrarily small error.
Using Newton-Schulz iteration of Pad\'{e} family \cite{Kennedy1991_sign}, we analytically obtain a parameter set that accurately builds QET of matrix sign functions.
Furthermore, the parameter set has constant unique values which do not depend on either an allowable error $\varepsilon$ or any parameter of the matrix.
Although our construction expenses the computational cost compared to the optimal protocol \cite{low2017-sign,Lin2020-ground_state} due to the strong limitation on the parameters, it suffers from no numerical instability and even can overwhelm the optimal one when we take recovery of coherent errors into account \cite{Tan2023-recovery_QSP}.
With various recursive constructions of matrix functions such as Newton iteration \cite{Higham2008-matrix} and logistic map \cite{Navickas2011-logistic,pawela2023-logistic}, r-QET/r-QSVT will give a promising candidate for executing complicated operations on quantum computers in accurate and stable ways.

\textit{Quantum Eigenvalue Transformation (QET).---} Throughout the main text, we concentrate on QET and thus r-QET for hermitian matrices for simplicity (See Supplemental Materials S2 for QSVT, whose discussion is completely parallel).
We begin with briefly introducing QET here.
Let a hermitian matrix $A$ have spectral decomposition $A = \sum_a a \ket{a}\bra{a}$ ($a \in \bbR$) on a finite-dimensional Hilbert space $\mcl{H}$.
Block-encoding of $A$ is defined by a unitary gate $O_A$ satisfying
\begin{equation}
    \braket{0|O_A|0}_b = \frac{A}{\alpha}, \quad \alpha > 0.
\end{equation}
Here, $\ket{0}_b$ denotes a reference state in an ancillary Hilbert space $\mcl{H}_b$.
We set $\alpha = 1$ by the renormalization $A \to A/\alpha$ below.
Construction of block-encoding is known for a linear combination of unitaries, a sparse-access matrix, and so on \cite{Low2019-qubitization}.

Combining parametrized unitary operations on the ancilla system,
\begin{equation}
    R_\phi = e^{i \phi (2\ket{0}\bra{0}_b - I_b)} \otimes I,
\end{equation}
we define a degree-$q$ QET operator by
\begin{eqnarray}
    && \mr{QET}[A,\vec{\phi}] = \nonumber \\
    && \quad \begin{cases}
    R_{\phi_1} O_A \prod_{i=1}^{(q-1)/2} \left[ R_{\phi_{2i}} O_A^\dagger R_{\phi_{2i+1}} O_A \right], & \text{$q$: odd}, \\
    \prod_{i=1}^{q/2} \left[ R_{\phi_{2i-1}} O_A^\dagger R_{\phi_{2i}} O_A \right], & \text{$q$: even}.
    \end{cases} \nonumber \\
    && \label{Eq:QET_definition}
\end{eqnarray}
By properly tuning the parameter set $\vec{\phi} \in \bbR^q$, it can realize various polynomial functions $f(x) = \sum_n c_n x^n$ ($c_n \in \bbC$) as
\begin{equation}\label{Eq:QET_block_encode}
    \braket{0|\mr{QET}[A,\vec{\phi}]|0}_b = f(A) = \sum_n c_n A^n.
\end{equation}
It is proven that there exists a parameter set $\vec{\phi}$ if and only if $f(x)$ satisfies all the following conditions \cite{Gilyen2019-qsvt}:
\begin{enumerate}[(i)]
    \item $f(x)$ has a degree at most $q$ and a parity $(-1)^q$.
    \item $|f(x)| \leq 1$ for any $x \in [-1,1]$ and $|f(x)| \geq 1$ for any $x \in (-\infty,-1] \cup [1,\infty)$.
    \item (If $q$ is even) $|f(ix)f^\ast(ix)| \geq 1$ for any $x \in \bbR$, where $f^\ast (x)$ is defined by $f^\ast(x)=\sum_n c_n^\ast x^n$.
\end{enumerate}
By $\order{1}$ controlled operations $\mr{QET}[A,\vec{\phi}]$, generic renormalized matrix functions $f(x)$ with $|f(x)| \leq 1/4$ ($^\forall x \in [-1,1]$) are also realizable.

For a desired function $f(x)$ satisfying (i)-(iii), how can we find a proper parameter set $\vec{\phi}$ ?
As far as we know, the analytical result is restricted to the Chebyshev polynomials $f(x)=T_n(x)$ with the trivial angles $\vec{\phi} = ((q-1)\pi/2,-\pi/2,-\pi/2,\hdots, - \pi/2)$, which can be utilized for Grover's search and its family such as amplitude amplification \cite{Grover1997-prl}.
In general, it requires finding all the roots of $1-f(x)f^\ast (x)$, which is a degree-$q$ polynomial in $x^2$, and iteratively decomposing $\mr{QET}[A,\vec{\phi}]$ into lower-degree QET operators \cite{Gilyen2019-qsvt}.
However, for useful functions such as $e^{-it A}$ (Hamiltonian simulation \cite{Low2017-qsp,Low2019-qubitization}), $A^{-1}/2 \kappa$ (quantum linear system problem, QLSP \cite{childs2017_inversion}), and $\sign (A)$ (eigenstate filtering \cite{Lin2020-ground_state}), the typical degree $q$ is quite large as $q \in \poly{N, \log (1/\varepsilon)}$ depending on the system size $N$ and the allowable error $\varepsilon$.
Thus, it is difficult to accurately compute $\vec{\phi}$ generally having $\poly{N, \log (1/\varepsilon)}$ different values in a numerically-stable way, although it can be done by $\poly{N, \log (1/\varepsilon)}$-time classical calculation.
This instability has been partially resolved by refining root-finding problems and decomposition into lower degrees \cite{Haah2019-numerical_QSP,Rui2020_numerical_QSP}, or employing optimization \cite{Dong2021-symmetric_QSP,Wang2022-symmetric_QSP}.
They have numerically succeeded up to $q \sim 10^4$ within $10^2 \sim 10^4$ seconds.

%
\textit{Recursive QET (r-QET) and Newton iteration.---} Here, we propose the protocol named recursive QET (r-QET), and provide the formulation combined with Newton iteration for matrix functions \cite{Higham2008-matrix}.
It aims to implement complicated matrix functions by QET with keeping tractability of the parameter set $\vec{\phi}$ in terms of analytical or numerically-cheap computation.
Our strategy is to employ recursive relations: while each step operation is executed in a simple way, its repetition forms rather complicated functions.

Suppose that we want to execute complicated matrix functions of $A$ with its block-encoding $O_A$.
Based on the fact that QET generates block-encoding from block-encoding as Eq. (\ref{Eq:QET_block_encode}), we organize r-QET by recursively defining a series of block-encodings $\{ O_{X_n} \}_n$ by
\begin{equation}\label{Eq:Recursive_block_encoding}
    O_{X_{n+1}} = \mr{QET}[O_{X_n},\vec{\phi}^g], \quad n = 0,1,2, \hdots.
\end{equation}
Here, the initial input $O_{X_0}$ is dependent on $O_A$ (e.g. $O_{X_0}=O_A$), and we have options for the parameter set $\vec{\phi}_g$ including its values and dimension.
With these options, the above construction forms a recursive relation of matrices $X_n = \braket{0|O_{X_n}|0}_b$,
\begin{equation}\label{Eq:Newton_iteration}
    X_{n+1} = g (X_n), \quad X_0 = \braket{0|O_{X_0}|0}_b,
\end{equation}
with a variety of polynomial functions $g$.
Recursive relations of matrices like Eq. (\ref{Eq:Newton_iteration}) are known to be available for complicated matrix functions exemplified by a matrix logistic map \cite{Navickas2011-logistic,pawela2023-logistic}.

One of the most promising candidates for the recursive relation is Newton iteration, which was originally invented for solving nonlinear equations \cite{Higham2008-matrix}.
Newton iteration for matrices enables us to efficiently compute various matrix functions $f(A)$ with properly choosing the function $g$ as a result of iteration $\lim_{N \to \infty} X_n = f(A)$.
For instance, $g(X) = (3 X - X^3)/2$ and $X_0 = A$ generates the matrix sign function $X_n \to \sign (A)$ (defined later) and the one by $g(X) = 2 X - X A X$ and $X_0 = \theta A$ ($0<\theta \ll 1$) generates the matrix inversion $X_n \to A^{-1}$ \cite{Higham2008-matrix}.
When r-QET combined with Newton iteration, we organize the parameter set $\vec{\phi}_g$ so that the corresponding polynomial function $g$ reproduces Newton iteration.
The iteration continues until it achieves an allowable error $\varepsilon$ as $\norm{X_n - f(A)} \leq \varepsilon$ ($\norm{\cdot}$ denotes the operator norm), and thus the iteration number $n$ depends on $\varepsilon$.
Then, the resulting unitary gate $O_{X_n}$ provides an accurate block-encoding for the function $f(A)$.

By iterative substitution of the recursive relation Eq. (\ref{Eq:Recursive_block_encoding}), the block-encoding $O_{X_n}$ is rewritten by
\begin{equation}\label{Eq:Repeated_QET_substitution}
    O_{X_n} = \mr{QET}[\mr{QET}[\hdots[\mr{QET}[O_{X_0},\vec{\phi}^g], \vec{\phi}^g],\hdots,\vec{\phi}^g],
\end{equation}
which has the form of $O_{X_n} = \mr{QET}[O_{X_0},\vec{\phi}_n]$.
The $(\deg (g^n))$-dimensional parameter set $\vec{\phi}_n$ is determined solely by the $(\deg (g))$-dimensional one $\vec{\phi}^g$.
Since the desirable error $\varepsilon$ solely affects the iteration number $n$, we can obtain a parameter set of QET achieving arbitrarily small error in an analytical or numerically-stable way.

While we concentrate on the usage of the standard QET for implementing the function $g$ here, we note that other types of QET/QSVT are also available.
For instance, if the recursive relation $g(X)$ include some matrices other than $X$ (e.g. $g(X)=2 X - XAX$ for matrix inversion), we need at-least multi-variate QET/QSVT \cite{Rossi_2022_multi_QSP,Weil_2023_multi_QSP}.

\begin{figure}[t]
    \includegraphics[height=4.3cm, width=9cm]{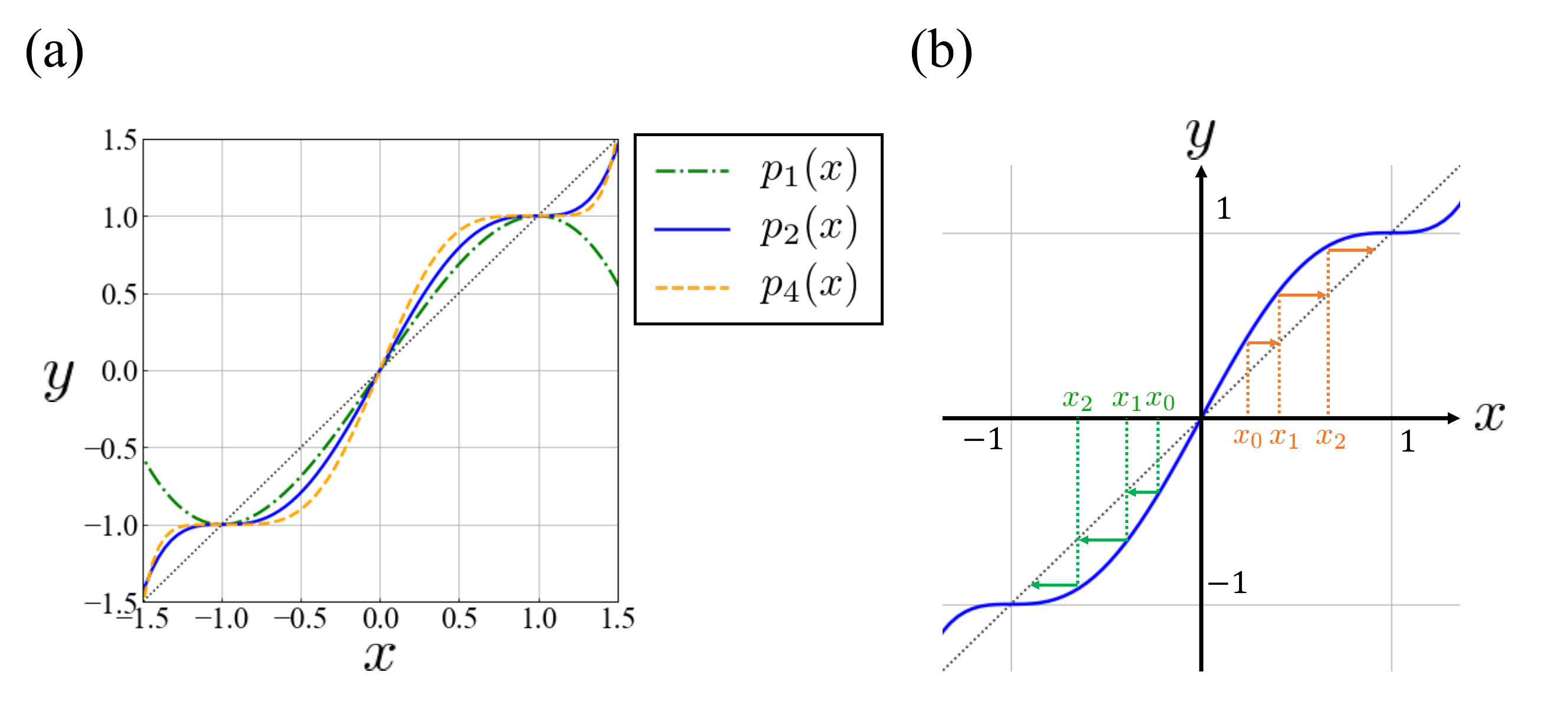}
    \caption{(a) Some members of Pad\'{e} family. (b) Intuitive picture of Newton iteration based on $p_2(x)$. Each eigenvalue of $X_n$, represented by $x_n$, approaches $1$ or $-1$ depending on its sign.} 
    \label{Figure}
\end{figure}

%
\textit{Analytical r-QET for Matrix Sign Function.---}
In r-QET, we should carefully choose the recursive relation so that the tractable function $g$ can be realized by QET and the resulting function $X_n$ is meaningful.
Here, we show the power of r-QET combined with Newton iteration by analytically constructing the parameter set of QET for matrix sign functions.

Suppose that a hermitian matrix $A = \sum_{a} a \ket{a}\bra{a}$ has a spectral gap $2\Delta$ ($> 0$) around zero as $a \in [-1, -\Delta] \cup [\Delta,1]$.
Then, the matrix sign function $\sign (A)$ is defined by
\begin{equation}
    \sign (A) = \sum_a \sign (a) \ket{a}\bra{a}, \quad \sign (a) = \frac{a}{|a|},
\end{equation}
which is useful for various tasks such as eigenstate filtering \cite{Lin2020-filtering_qlsp}.
The matrix sign function $\sign (A)$ can be generated by Newton-Shultz iteration using a series of rational functions called Pad\'{e} family $\{ p_l (x) \}_{l \in \bbN}$ with the initial input $X_0 = A$ \cite{Higham2008-matrix,Kennedy1991_sign}.
As the recursive relation $g(X)$, we adopt the second simplest case,
\begin{equation}\label{Eq:Pade_2}
    p_2(X) = \frac18 (15 X - 10 X^3 + 3 X^5), \quad \deg (p_2) = 5,
\end{equation}
since the simplest one $p_1(X)=(3X-X^3)/2$ violates the condition (ii) required for QET [See Fig. \ref{Figure} (a)].
The convergence to $\sign (A)$ is confirmed by the fact that every eigenvalue of $X_n$, denoted by $x_n$, moves to $+1$ or $-1$ based on the recursive relation $x_{n+1} = p_2 (x_n)$ [See Fig. \ref{Figure} (b)].

It is sufficient to determine the parameter set $\vec{\phi}^{p_2} \in \bbR^5$, which cosntructs the relation,
\begin{equation}\label{Eq:recursive_QET_Pade_2}
    O_{X_{n+1}} = \mr{QET}[O_{X_n},\vec{\phi}^{p_2}], \quad O_{X_0} = O_A.
\end{equation}
Finding $\vec{\phi}^q$ can be attributed to finding all the roots of $1-p_2(x)p_2^\ast(x)$.
While it is generally hard in an analytical way when the degree is larger than $5$, we find them by the factorization,
\begin{equation}\label{Eq:Factorization_Pade_2}
    1-p_2(x)p_2^\ast(x) = -\frac{9}{64} (1-x^2)^3 (x^2-s)(x^2-s^\ast),
\end{equation}
with $s = (11+3\sqrt{15}i)/6$.
This leads to the following analytical parameter set $\vec{\phi}^{p_2}$ (See Supplementary Materials S1 for its detailed calculation):
\begin{eqnarray}
    && \phi_1^{p_2} = 0, \quad \phi_2^{p_2} = \pi + \frac12 \arctan \frac{\sqrt{15}}7, \nonumber \\
    && \phi_3^{p_2} = \pi + \frac12 \arctan \sqrt{15}, \quad \phi_4^{p_2} = - \frac12 \arctan \sqrt{15}, \nonumber \\
    && \phi_5^{p_2}= - \frac12 \arctan \frac{\sqrt{15}}7.
    \label{Eq:Parameter_set_Pade_2}
\end{eqnarray}
Therefore, by repeated substitution Eq. (\ref{Eq:Repeated_QET_substitution}), we obtain an analytical QET operator $\mr{QET}[O_A,\vec{\phi}_n]$ for the matrix sign function $\sign (A)$. 
The parameter set $\vec{\phi}_n$ has only eight different angles $\pm \phi_i^{p_2}$ for $i=2,3,4,5$, and they always appear in the fixed orders of $\phi_2^{p_2} \to \phi_3^{p_2} \to \phi_4^{p_2} \to \phi_5^{p_2}$ or $-\phi_5^{p_2} \to - \phi_4^{p_2} \to - \phi_3^{p_2} \to - \phi_2^{p_2}$.

\textit{Cost for Matrix Sign Function.---}
Let us evaluate the cost for the matrix sign function.
We repeat the recursive relation until the desirable error $\varepsilon \in [0,1]$ is achieved as $\norm{X_n - \sign (A)} \leq \varepsilon$.
The convergence rate to $\sign (A)$ based on the Newton iteration, Eq. (\ref{Eq:Pade_2}), is dominated by the gap $\Delta$ as follows \cite{Kennedy1991_sign}:
\begin{eqnarray}\label{Eq:Error_Pade_2}
    \norm{X_n - \sign (A)} \leq (1-\Delta^2)^{3^n}.
\end{eqnarray}
By using the relation $\log (1-\Delta^2)^{-1} \geq \Delta^2$ for $\Delta \in [0,1)$, it is sufficient to choose the iteration number by
\begin{equation}\label{Eq:Iteration_Pade_2}
    n = \left\lceil \log_3 \left( \frac{1}{\Delta^2} \log (1/\varepsilon) \right) \right\rceil.
\end{equation}

The cost is measured by the query complexity, i,e., the number of the oracle $O_A$ in the unitary gate $O_{X_n}$ giving $X_n = \sign (A) + \order{\varepsilon}$.
Let the query complexity of $O_{X_n}$ denote $q_n$.
Then, the recursive relation Eq. (\ref{Eq:recursive_QET_Pade_2}) immediately implies $q_{n+1} = 5 q_n$ and $q_0 = 1$.
Under the proper iteration number $n$ by Eq. (\ref{Eq:Iteration_Pade_2}), we need the query complexity given by
\begin{eqnarray}
    q_n &=& 5^n \nonumber \\
    &\leq& 5 \times  \left( \frac{1}{\Delta^2} \log (1/\varepsilon) \right)^{\log_3 5} \\
    &\in& \Theta \left( \frac{1}{\Delta^{2\log_3 5}} \log^{\log_3 5} (1/\varepsilon) \right). \label{Eq:Query_complexity_Pade_2}
\end{eqnarray}

Let us compare with the standard QET approach \cite{low2017-sign,Gilyen2019-qsvt,Martyn2021-grand-unif}, which uses polynomial approximation of the error function $\mr{erf}(kx) \simeq \sign (x)$ with large $k >0$.
This yields the query complexity $\Theta \left( \Delta^{-1} \log (1/\varepsilon) \right)$ to implement $\sign (A) + \order{\varepsilon}$, which is known to be optimal both in $\Delta$ and $\varepsilon$.
Considering the value $\log_3 5 \simeq 1.465$, the query complexity of r-QET, Eq. (\ref{Eq:Query_complexity_Pade_2}), is polynomially larger than the optimal one.
This difference comes from the flexibility of the parameter set $\vec{\phi}$.
The standard QET approach uses $\order{\Delta^{-1}\log(1/\varepsilon)}$ different parameters $\vec{\phi}$ obtained by numerically solving an $\order{\Delta^{-1}\log(1/\varepsilon)}$-degree equation.
In contrast, our approach employs only eight different values in a fixed order independent of $\Delta$ and $\varepsilon$.
At some expense of the query complexity, r-QET can serve $\sign (A)$ up to arbitrarily small error $\varepsilon$, with completely avoiding numerical instability and working only with a few kinds of gates.

While r-QET fails to achieve the optimal query complexity, it can overwhelm the optimal protocol when correcting coherent error is taken into account.
In QET, the multiplicative coherent error on the parameter set $\phi_i \to \phi_i (1+\delta)$ ($i=1,2,\hdots,q$) is a possible obstacle to accurate implementation of matrix functions.
Recently, A. K. Tan, et al. \cite{Tan2023-recovery_QSP} have shown that the coherent error $\delta$ on degree-$q$ QET can be suppressed up to $\order{\delta^{k+1}}$ ($k \in \bbN$) by additional query complexity,
\begin{equation}
    q_\mr{correct} \in \order{2^k (c_{\vec{\phi}})^{k^2} q}.
\end{equation}
Here, $c_{\vec{\phi}}$ denotes the number of different values in $\vec{\phi} \in \bbR^q$.
The standard QET approach achieving the optimality is expected to have $c_{\vec{\phi}} \in \order{q}$ different values in $\vec{\phi}$, and thereby its total query complexity amounts to $\mathcal{O}(\Delta^{-k^2-1} \log^{k^2+1} (1/\varepsilon))$.
By contrast, r-QET employs constant values with $c_{\vec{\phi}}=8$ regardless of any other parameters.
Therefore, the total query complexity including the recovering remains Eq. (\ref{Eq:Query_complexity_Pade_2}) as long as $k \in \order{1}$, and it can overwhelm the originally-optimal protocol for arbitrary $k \geq 2$.

\textit{Generalization of r-QET for matrix sign functions.---} 
Our result for matrix sign functions can be generalized to other members of the Pad\'{e} family or implementation of polar decomposition by r-QSVT.

In the first case, the recursive relation $X_{n+1}=p_{l}(X_n)$ with $X_0 = A$ ($l \in \bbN$) by the Pad\'{e} family, 
\begin{equation}
    p_l(x) = x \sum_{k=0}^l \frac{(2k-1)!!}{2^k k!} (1-x^2)^k.
\end{equation}
casts the matrix sign function as $\lim_{n \to \infty} X_n = \sign (A)$.
The conditions (i)-(iii) for QET can be satisfied only when $l$ is even \cite{footnote_Pade_even}.
r-QET requires only $2l$ different values for the parameter set for the degree-$(2l+1)$ polynomial $p_l(X)$, where one of them can be zero.
As well as $l=2$, we can analytically determine the parameter set for $l=4$ by the quartic formula.
Even for larger $l \geq 6$, the numerical instability for the parameter determination is much more suppressed than the standard QET.
The advantage of generalization to larger degrees is the computational cost.
The convergence rate to $\sign (A)$ becomes faster as $\norm{X_n - \sign (A)} \leq (1-\Delta^2)^{(l + 1)^n}$ \cite{Kennedy1991_sign}, and hence the query complexity $q_n$ in $O_A$ becomes smaller as
\begin{eqnarray}
    && q \in \Theta \left( \frac{1}{\Delta^{2(1+\nu_{l})}} \log^{1+\nu_{l}} (1/\varepsilon)\right), \label{Eq:Query_complexity_generic_Pade}\\
    && \nu_{l} = \frac{\log (2l +1)}{\log (l + 1)} - 1.
\end{eqnarray}
r-QET achieves the scaling $\Theta (\Delta^{-(2+o(1))} \log^{1+o(1)} (1/\varepsilon))$ for large $l$ \cite{footnote_Pade_even}.
While it is still not optimal in the gap $\Delta$, it can reach the optimal one in the desirable error $\varepsilon$.

Let us discuss the generalization to r-QSVT.  
QSVT produces block-encoding executing polynomial transformation of every singular value from block-encoding of a generic matrix $A$, and hence r-QSVT can be composed by recursive iteration of QSVT (See Supplemental Materials S2 in detail).
When we use the same parameter set $\vec{\phi}^{p_2}$ for the matrix sign function, r-QSVT organizes block-encoding $\{ O_{X_n} \}_n$ reproducing 
\begin{equation}
    X_{n+1} = \frac{X_n}8 \left\{ 15 - 10 X_n^\dagger X_n + 3 (X_n^\dagger X_n)^2 \right\}, \quad X_0 = A.
\end{equation}
Under this recursive iteration, the matrix converges to $X_n \to A (\sqrt{A^\dagger A})^{-1}$, which is the unitary part of the polar decomposition of the non-singular matrix $A$.
With the same iteration number Eq. (\ref{Eq:Iteration_Pade_2}), r-QSVT achieves an arbitrarily small error $\varepsilon$ and analyticity (or numerical stability) of the parameter set for the polar decomposition.

\textit{Discussion and Conclusion.---} In this Letter, we propose recursive QET/QSVT that executes recursive relations by QET/QSVT.
All the parameters can be determined by low-degree polynomials for recursive relations, which enables analytical or numerically-stable calculation and also leads to feasibility of recovering coherent noise.
Particularly, the construction of matrix sign function when combined with Newton iteration is the first analytical result on parameters for useful functions other than the trivial Chebyshev polynomials.
Indeed, with the analytically-obtained parameters given by Eq. (\ref{Eq:Parameter_set_Pade_2}), we can execute eigenstate filtering, and thereby solve quantum linear system problems \cite{Lin2020-filtering_qlsp} (See also Supplemental Materials S3).
It will not only contribute to understanding the expressibility of QET with restricted parameters \cite{Dong2021-symmetric_QSP,Wang2022-symmetric_QSP}, but also open up an efficient path to construct QET for various purposes.

We conclude this Letter with some future directions of r-QET/r-QSVT.
The first direction is to explore accurate implementation of other matrix functions by Newton iteration.
We have concentrated on using the standard QET as a subroutine for the matrix sign function for simplicity.
However, Newton iteration covers various matrix functions and can show quadratic or faster convergence.
Exploiting a series of QET/QSVT protocols, such as those for Fourier series expansions \cite{Dong_2022_PRXQuantum_Fourier_QSP,Silva_2022_Fourier_QSP,Xin_2022_Fourier_QSP} and multivariate polynomial functions \cite{Rossi_2022_multi_QSP,Weil_2023_multi_QSP}, to reproduce their recursive relations will provide alternative ways to execute various matrix functions with efficiently and accurately providing the required parameters.

Second, it will be also important to seek for the usage of various recursive relations.
For instance, recursive relations based on the matrix logistic map can generate either chaotic or non-chaotic behaviors of each eigenvalue \cite{Navickas2011-logistic,pawela2023-logistic}.
While its usefulness in quantum many-body systems is still controversial, examining compatibility of such recursive relations with r-QET/r-QSVT will open up a new way of quantum operations by QET/QSVT.

\begin{acknowledgments}
K. M. is supported by RIKEN Special Postdoctoral Researcher Program.
This work is supported by MEXT Quantum Leap Flagship Program (MEXTQLEAP) Grant No. JPMXS0118067394, JPMXS0120319794, and JST COI-NEXT program Grant No. JPMJPF2014.
\end{acknowledgments}

\bibliography{bibliography.bib}

\begin{thebibliography}{26}
\bibitem [{foo()}]{footnote_Pade_even}%
  \BibitemOpen
  \href@noop {} {}\bibinfo {note} {Exactly speaking, we was not able to found
  the proof of the latter part of the condition (ii) for generic even $l$,
  $|p_l(x)| \geq 1$ for any $x \notin [-1,1]$, although it seems to be correct
  from their graphs. By computing the derivatives of $p_l (x)$ at $x= \pm 1$,
  we have proven that $p_l(x)$ is compatible with the standard QET at least up
  to $l=10^3$. While we believe that the exponent $\nu_l$ [See Eq.
  (\ref{Eq:Query_complexity_generic_Pade})] can be arbitrarily small under $l
  \to \infty$, it is ensured to reach $\nu_{10^3} \simeq 0.100$ at the present
  stage.}\BibitemShut {Stop}%
\end{thebibliography}

\clearpage

\renewcommand{\thesection}{S\arabic{section}}
\renewcommand{\theequation}{S\arabic{equation}}
\setcounter{equation}{0}
\renewcommand{\thefigure}{S\arabic{figure}}
\setcounter{figure}{0}

\onecolumngrid
\begin{center}
 {\large 
 {\bfseries Supplemental Materials for \\ ``Recursive Quantum Eigenvalue/Singular-Value Transformation; \\ Analytic Construction of Matrix Sign Function by Newton iteration'' }}
\end{center}

\begin{center}
Kaoru Mizuta$^{1,2}$ and Keisuke Fujii$^{3,4,1,5}$
\end{center}

\begin{center}
{\small 
\textit{$^1$RIKEN Center for Quantum Computing (RQC), Hirosawa 2-1, Wako, Saitama 351-0198, Japan} \\
\textit{$^2$Department of Applied Physics, The University of Tokyo, Hongo 7-3-1, Bunkyo, Tokyo 113-8656, Japan} \\
\textit{$^3$Graduate School of Engineering Science, Osaka University, \\ 1-3 Machikaneyama, Toyonaka, Osaka 560-8531, Japan.} \\
\textit{$^4$Center for Quantum Information and Quantum Biology, Osaka University, Japan.} \\
\textit{$^5$Fujitsu Quantum Computing Joint Research Division at QIQB,
\\ Osaka University, 1-2 Machikaneyama, Toyonaka 560-0043, Japan}
\\ (Dated: \today)
}
\end{center}
\vspace{20pt}

\twocolumngrid

\section{Parameter determination of QET}\label{Sec_supp:Parameter_determination}
In this section, we show how to determine the parameter set $\vec{\phi} \in \bbR^q$ in QET, for this Letter to be self-contained.
The discussion here is mainly based on Ref. \cite{Gilyen2019-qsvt}.

Suppose that we want to realize a degree-$q$ polynomial $f(x)$ satisfying (i)-(iii) in the main text by degree-$q$ QET.
This exploits a technique of quantum signal processing (QSP) \cite{Low2017-qsp}, since the achievable functions are equivalent to QET/QSVT by so-called qubitization technique \cite{Low2019-qubitization}.
We begin with determining the parameter set for QSP to realize the function $f(x)$.
QSP is a repetition of single-qubit rotation operators that embodies a polynomial function of a certain signal $x \in [-1,1]$.
A rotation-based degree-$q$ QSP operator $\mr{QSP}[x,\vec{\phi}]$ on $\bbC^2$ is defined by
\begin{eqnarray}
    \mr{QSP}[x,\vec{\phi}^\prime] &=& e^{i \phi_0^\prime Z} \prod_{i=1}^q \left[ W(x) e^{i \phi_i^\prime Z}\right], \\
    W(x) &=& \left( \begin{array}{cc}
        x & i \sqrt{1-x^2} \\
        i \sqrt{1-x^2} & x
    \end{array} \right),
\end{eqnarray}
where $\vec{\phi}^\prime \in \bbR^{q+1}$ is a tunable parameter set and $Z$ is the Pauli-$Z$ operator.
The QSP operator can be written in the form of
\begin{equation}
    \mr{QSP}[x,\vec{\phi}^\prime] = \left( \begin{array}{cc}
        f(x) & i h (x) \sqrt{1-x^2} \\
        i h^\ast (x) \sqrt{1-x^2} & f^\ast (x),
    \end{array}\right)
\end{equation}
and there exists a corresponding parameter set $\vec{\phi}^\prime \in \bbR^{q+1}$ if and only if $f(x)$ satisfies (i)-(iii) in the main text.
Then, a polynomial function $h(x)$ satisfies
\begin{enumerate}[(i$^\prime$)]
    \item $h(x)$ has a degree at most $q-1$
    \item $h(x)$ has a parity $(-1)^{q-1}$.
    \item $f(x)f^\ast(x)+(1-x^2) h(x) h^\ast (x) = 1$ for any $x$.
\end{enumerate}

To obtain a parameter set $\vec{\phi}^\prime$ for the rotation-based QSP, we should obtain the polynomial function $h(x)$.
This can be executed by finding all the roots $\{s_i \}_{i=1}^q$ of $1-f^\ast (x) f(x)$, which is a degree-$q$ polynomial in $x^2$.
From the conditions (i)-(iii) in the main text, the polynomial $1-f^\ast (x) f(x)$ is always factorized in the form of
\begin{equation}\label{Seq:Root_finding}
    1- f^\ast(x) f(x) = \alpha (1-x^2) \prod_{s_i; \mr{Im}(s_i) \geq 0} (x^2 - s_i)(x^2-s_i^\ast)
\end{equation}
with $\alpha > 0$, and thus 
\begin{equation}
    h(x) = \sqrt{\alpha} \prod_{s_i \in S; \mr{Im}(s_i) \geq 0} (x^2 - s_i),
\end{equation}
gives a proper choice satisfying (i$^\prime$)-(iii$^\prime$).

Once one obtains $h(x) = \sum_{n=0}^{q-1} h_n x^n$ for the desired polynomial $f(x) = \sum_{n=0}^q f_n x^n$, the parameter set $\vec{\phi}^\prime$ is determined in a recursive way.
We define $\phi_q^\prime$ by
\begin{equation}\label{Seq:Angle_finding}
    e^{2i \phi_q^\prime} = \frac{f_q}{h_{q-1}},
\end{equation}
and then we update the polynomials $f(x)$ and $h(x)$ by
\begin{eqnarray}
       \tilde{f}(x) &=& e^{-i \phi_q^\prime} x f(x) + e^{i \phi_q^\prime} (1-x^2) h(x) \label{Seq:degrade_fx},\\
       \tilde{h}(x) &=& e^{i \phi_q^\prime} x h(x) - e^{-i \phi_q^\prime} f(x). \label{Seq:degrade_hx}
\end{eqnarray}
The pair of $\tilde{f}(x)$ and $\tilde{h}(x)$ satisfies (i)-(iii) and (i$^\prime$)-(iii$^\prime$) for the degree $q-1$, and hence it can be realized by a degree-$(q-1)$ QSP.
The parameter $\phi_q^\prime$ is determined by Eq. (\ref{Seq:Angle_finding}) with using $\tilde{f}(x)$ and $\tilde{h}(x)$ instead of $f(x)$ and $h(x)$.
We repeat this procedure until the obtained polynomials become trivial.

Finally, we transform the angles $\vec{\phi}^\prime$ into those for QET.
With a proper choice of two-dimensional basis, QET can be identified with a reflection-based QSP operation $\bbC^2$, which is defined by
\begin{eqnarray}
    \overline{\mr{QSP}}[x,\vec{\phi}] &=& \prod_{i=1}^q \left[ e^{i \phi_i Z} R(x)\right], \quad \vec{\phi} \in \bbR^q, \\
    R(x) &=& \left( \begin{array}{cc}
        x & \sqrt{1-x^2} \\
        \sqrt{1-x^2} & -x
    \end{array}\right).
\end{eqnarray}
Namely, QET and reflection-based QSP realize the same function by $\braket{0|\mr{QET}[A,\vec{\phi}]|0}$ and $\braket{0|\overline{\mr{QSP}}[x,\vec{\phi}]|0}$ with the same angles $\vec{\phi}$.
Achievable functions by them are equivalent to those by rotation-based QSP, where we choose the angles by
\begin{equation}
    \phi_1 = \phi_0^\prime + \phi_q^\prime + (q-1)\frac{\pi}2, \quad \phi_i = \phi_{i-1}^\prime - \frac{\pi}2 \quad (2 \leq i \leq q).
\end{equation}

Numerical instability arises in the root-finding problem of the degree-$q$ polynomial, Eq. (\ref{Seq:Root_finding}), and the error is accumulated via iterative computation by Eqs. (\ref{Seq:Angle_finding})-(\ref{Seq:degrade_hx}).
In Newton-iteration-based QET, we do not suffer from them since we employ low-degree QET.
When we consider $f(x)=p_2(x)=(15x-10x^3+3x^3)/8$ to construct a matrix sign function as the main text, the factorization Eq. (\ref{Eq:Factorization_Pade_2}) analytically gives the pairwise function $h(x)$ by
\begin{equation}
    h(x) = \frac{3}{8} x^4 - \frac{1}{16} (17+3\sqrt{15}i) x^2 + \frac{1}{16}(11+3\sqrt{15}i).
\end{equation}
The iterative determination of the angles by Eqs. (\ref{Seq:Angle_finding})-(\ref{Seq:degrade_hx}) can be analytically completed, which results in Eq. (\ref{Eq:Parameter_set_Pade_2}).
We ensure the analyticity of the angles $\vec{\phi}$ also when we use the higher-order Pad\'{e} family $f(x)=p_4(x)$, where $h(x)$ can be analytically computed by the factorization,
\begin{eqnarray}
    && 1- (p_4(x))^2 = -(1-x^2)^5 \sum_{i=0}^4 a_i (x^2)^i, \\
    && a_0 = 1, \quad a_1 = -\frac{17305}{16384}, \quad a_2 = \frac{14235}{16384}, \\
    && a_3 = - \frac{6475}{16384}, \quad a_4 = \frac{1225}{16384},
\end{eqnarray}
and the quartic formula.

\section{Extension to QSVT}\label{Sec_supp:qsvt}

In this section, we briefly discuss the extension to r-QSVT, especially for the matrix sign function.
Construction similar to the main text results in QSVT with tractable parameters for polar decomposition of generic matrices.

We begin with introducing QSVT \cite{Gilyen2019-qsvt}, with restricting to odd degrees for simplicity.
For a generic finite-dimensional matrix $A \in \bbC^{d_1 \times d_2}$, its block-encoding is defined by a unitary operator $O_A$ such that
\begin{equation}
    \Pi_1 O_A \Pi_2 = A
\end{equation}
is satisfied with certain $d_1$- and $d_2$-dimensional projections $\Pi_1$ and $\Pi_2$.
With parametrized rotations designated by
\begin{equation}
    R_\phi^{\Pi_1} = e^{i\phi (2\Pi_1 - I)}, \quad R_\phi^{\Pi_2} = e^{i\phi (2\Pi_2 - I)},
\end{equation}
a degree-$q$ QSVT operator for odd $q \in \bbN$ with a parameter set $\vec{\phi} \in \bbR^q$ is defined by
\begin{equation}
    \mr{QSVT}[O_A,\vec{\phi}] = R_{\phi_1}^{\Pi_1} O_A \prod_{i=1}^{(q-1)/2} \left[ R_{\phi_{2i}}^{\Pi_2} O_A^\dagger R_{\phi_{2i+1}}^{\Pi_1} O_A \right].
\end{equation}
The matrix $A$ is always described by singular value decomposition (SVD) as $A = \sum_\sigma \sigma \ket{u_\sigma}\bra{v_\sigma}$, where $\sigma$ ($> 0$), $\ket{u_\sigma} \in \bbC^{d_1}$, and $\ket{v_\sigma} \in \bbC^{d_2}$ are respectively called a singular value, a left singular vector, and a right singular vector.
QSVT executes polynomial processing of every singular value given by
\begin{equation}
    \Pi_1 \mr{QSVT}[O_A,\vec{\phi}] \Pi_2 = \sum_\sigma f(\sigma) \ket{u_\sigma}\bra{v_\sigma},
\end{equation}
where the scalar function $f(x)$ satisfies all the conditions (i)-(iii) in the main text.
We note that this is not equal to a polynomial function $f(A)$ in general, unlike QET.

Recursive QSVT (r-QSVT) is organized completely in a similar manner to r-QET; We construct a series of unitaries $\{ O_{X_n} \}_n$ by
\begin{equation}\label{Seq:recursive_QSVT}
    O_{X_{n+1}} = \mr{QSVT}[O_{X_n},\vec{\phi}^g],
\end{equation}
which begins from a unitary $O_{X_0}$ such that $\Pi_1 O_{X_0} \Pi_2 = X_0$.
It casts a recursive relation as
\begin{eqnarray}
    X_{n+1} &=& \Pi_1 O_{X_{n+1}} \Pi_2 \nonumber \\
    &=& \sum_{\sigma_n} g(\sigma_n) \ket{u_{\sigma_n}}\bra{v_{\sigma_n}},
\end{eqnarray}
where SVD of $X_n$ is given by $X_n = \sum_{\sigma_n} \sigma_n \ket{u_{\sigma_n}} \bra{v_{\sigma_n}}$.
If we choose the function $g$ based on Newton iteration, some complicated matrix functions of $A$ will be realized.
For instance, when the parameter set $\vec{\phi}^g$ is chosen as Eq. (\ref{Eq:Pade_2}), which is given for the matrix sign function, it forms the recursive relation,
\begin{eqnarray}
    X_{n+1} &=& \sum_{\sigma_n} \frac18 (15 \sigma_n - 10 \sigma_n^3 + 3 \sigma_n^5)\ket{u_{\sigma_n}}\bra{v_{\sigma_n}} \nonumber \\
    &=& \frac18 \left\{15 - 10 (X_n X_n^\dagger) + 3 (X_n X_n^\dagger)^2 \right\} X_n, \nonumber \\
    &&
\end{eqnarray}
with $X_0 = A$.
This is nothing but the one for computing the polar decomposition $U_A \in \bbC^{d_1 \times d_2}$, which is a unitary such that $A = U_A P_A$ with a positive semidefinite matrix $P_A \in \bbC^{d_2 \times d_2}$.
Namely, we obtain $\Pi_1 O_{X_n} \Pi_2 \to U_A$ by the recursive construction Eq. (\ref{Seq:recursive_QSVT}).
Since the convergence rate is the same as the one of matrix sign functions under the assumption $^\forall \sigma \in [\Delta,1]$, the query complexity to $O_A$ to achieve an allowable error $\varepsilon$ is given by Eq. (\ref{Eq:Query_complexity_Pade_2}).

We can see the realization of polar decomposition from another aspect.
The components of the polar decomposition are respectively represented by
\begin{equation}\label{Seq:Polar_decomposition_SVD}
    U_A = \sum_\sigma \ket{u_\sigma}\bra{v_\sigma}, \quad P_A = \sum_\sigma \sigma \ket{v_\sigma}\bra{v_\sigma}.
\end{equation}
Let us consider a QSVT operator $\mr{QSVT}[O_A,\vec{\phi}]$ with organizing $\vec{\phi}$ from $\vec{\phi}^g$ as well as the matrix sign function.
When the iteration number $n$ is large enough to satisfy Eq. (\ref{Eq:Iteration_Pade_2}), QSVT processes every singular value by $f(\sigma) = \sign (\sigma) + \order{\varepsilon} = 1 + \order{\varepsilon}$.
As a result, the QSVT operator $\mr{QSVT}[O_A,\vec{\phi}]$ can accurately reproduce the polar decomposition $U_A$ given by the formula Eq. (\ref{Seq:Polar_decomposition_SVD}).
In a similar way, r-QSVT with other even-order members of Pad\'{e} family $g_{2l^\prime} (x)$ outputs the polar decomposition of generic matrices, with yielding the query complexity Eq. (\ref{Eq:Query_complexity_generic_Pade}).

\section{Quantum algorithms with \\ matrix sign function of r-QET} \label{Sec_supp:application}

In the main text, we have shown that r-QET/r-QSVT provides analytical parameter sets for matrix sign functions.
Here, we provide that the same parameter sets are available for eigenstate filtering, eigenstate preparation, and quantum linear system problems (QLSP) as a complementary result by combining the results of Ref. \cite{Lin2020-filtering_qlsp}.

\subsection{Eigenstate filtering and eigenstate preparation}
Eigenstate filtering is to filter out eigenstates having eigenenergies larger than a certain cutoff.
Assuming that a hermitian matrix $A= \sum_{a \neq 0, |a| \leq 1} a \ket{a}\bra{a}$ has a spectral gap at $a=0$, it is equivalent to apply the projection,
\begin{equation}
     P_+(A) = \frac{1+\sign (A)}{2},
\end{equation}
We use the matrix sign function generated by r-QET, such that $\braket{0|O_{X_n}|0}_b = \sign (A) + \order{\varepsilon}$.
Then, the controlled operation defined by
\begin{eqnarray}
    && O_{\Delta} (A) = \nonumber \\
    && \quad e^{i (\pi/4) Y_{b^\prime}}\left( \ket{0}\bra{0}_{b^\prime} \otimes I + \ket{1}\bra{1}_{b^\prime} \otimes O_{X_n} \right) e^{i (\pi/4) Y_{b^\prime}}. \nonumber \\
    && \label{Seq:block_encode_filtering}
\end{eqnarray}
reproduces the filtering with additional one qubit $b^\prime$ by $\braket{00|O_\Delta (A)|00}_{bb^\prime} = P_+(A) + \order{\varepsilon}$.
The query complexity in $O_A$ for the eigenstate filtering is
\begin{equation}
    \order{\Delta^{-2(1+\nu_l)} \log^{1+\nu_l} (1/\varepsilon)},
\end{equation}
when we use the matrix sign function by r-QET with the function $p_l(x)$ [See Eq. (\ref{Eq:Query_complexity_generic_Pade})].
While it is worse than the optimal cost $\order{\Delta^{-1} \log (1/\varepsilon)}$, achieved by the standard QET \cite{Lin2020-filtering_qlsp}, r-QET determines all the parameters composed of $4l+1$ different rotation angles in numerically-cheaper ways.
For instance, it requires only $9$ different rotational angles, which are analytically given by $\{\pm \phi_i\}_{i=2,3,4,5}$ in Eqs. (\ref{Eq:Parameter_set_Pade_2}) and $\pi/4$ in Eq. (\ref{Seq:block_encode_filtering}), in the simplest case $l=2$.

Eigenstate preparation is to prepare a certain isolated eigenstate of $A$ (here, we set the eigenvalue at zero).
Assuming that $A$ has no eigenvalue in $[-\Delta,\Delta]$ except for the zero eigenvalue, it is completed by applying the projection,
\begin{equation}
    P_0(A) = \ket{0}\bra{0} = P_+ \left( \frac{A+\Delta /2}{1+\Delta /2} \right) P_+ \left( - \frac{A-\Delta/2}{1+\Delta/2} \right),
\end{equation}
to a proper initial state $\ket{\psi_0}$.
In the above filtering, we use the block-encoding of $(A \pm \Delta/2)/(1+\Delta/2)$ for normalization, which requires one query to a controlled-$O_A$ operation.
We also need single-qubit unitary gates on an additional qubit, whose rotation angle is $\pm \arctan \sqrt{\Delta/2}$.
When the initial state $\ket{\psi_0}$, prepared by a unitary $U_0$, has the overlap $\gamma = |\braket{0|\psi_0}|$, r-QET with $p_l(x)$ requires
\begin{equation}
    \order{\frac{1}{\gamma} \Delta^{-2(1+\nu_l)} \log^{1+\nu_l} (1/\gamma \varepsilon)}
\end{equation}
queries to $O_A$ and $\order{\gamma^{-1}}$ queries to $U_0$.
Here, we use the amplitude amplification \cite{Martyn2021-grand-unif} which allows the quadratic speedup from the time $\order{\gamma^{-2}}$, expected by the success probability.
At the expense of the cost to some extent compared to the optimal one $\order{\gamma^{-1} \Delta^{-1}\log(1/\varepsilon)}$ \cite{Lin2020-filtering_qlsp}, r-QET works only with $4l+3$ different rotation angles.
In addition, they can be analytically obtained in the simplest cases $l=2,4$, or avoid numerical instability for larger $l$.

\subsection{Quantum linear system problem (QLSP)}
Quantum linear system problem (QLSP) is to obtain $\ket{x}=A^{-1}\ket{b}$ from a given non-singular hermitian matrix $A$ and a given vector $\ket{b}$.
Here, we define the condition number of $A$ by $\kappa = \Delta^{-1}$, and the vector $\ket{b}$ is prepared by a unitary gate $U_b$.
Recently, it has been shown that QLSP can be solved by using eigenstate filtering and eigenstate preparation as subroutines \cite{Lin2020-filtering_qlsp}.
This relies on the fact that $\ket{x}$ is obtained by an isolated eigenstate as
\begin{eqnarray}
    && \tilde{A} \left( \begin{array}{c}
        \ket{x} \\
        0
    \end{array}\right)= 0, \\
    && \tilde{A} = \left( \begin{array}{cc}
        0 & A(I-\ket{b}\bra{b}) \\
        (I-\ket{b}\bra{b}) A  & 0
    \end{array}\right).
\end{eqnarray}
The block-encoding of $\tilde{A}$ is organize by $\order{1}$ queries to $O_A$ and $U_b$.
The algorithm prepares a state having an $\order{1}$ overlap with $^\mr{t}(\ket{x},0)$ by Zeno effect reproduced by eigenstate filtering, and then make a projection onto $^\mr{t}(\ket{x},0)$ within $\order{\varepsilon}$ by eigenstate preparation in $\tilde{A}$.

In the first step, the algorithm begins with preparing the zero-energy eigenstate $^\mr{t}(\ket{b},0)$ of the trivial Hamiltonian $\tilde{A}_0 = X \otimes (I-\ket{b}\bra{b})$ 
Then, to approximately obtain the eigenstate of the target Hamiltonian $\tilde{A}$, it uses a path
\begin{equation}
    \tilde{A}(s) = f_s \tilde{A} + (1-f_s) \tilde{A}_0,
\end{equation}
with a specific scheduling function $f_s$ satisfying $f_{s=0}=0$ and $f_{s=1}=1$.
The algorithm mimics Zeno effect along this path by a series of eigenstate filtering, which is represented by
\begin{equation}
    \prod_{t=0}^{M} P_0 \left( \tilde{A}(t/M) \right)
    \left( \begin{array}{c}
        \ket{b} \\
        0
    \end{array}\right)
    = c  \left( \begin{array}{c}
        \ket{x} \\
        0
    \end{array}\right) + \dots.
\end{equation}
To achieve the overlap $c \in \order{1}$ with $^\mr{t}(\ket{x},0)$, we need $M \in \order{(\log \kappa)^2}$-times repetition and the error smaller than $\varepsilon_\mr{EF} \in \order{M^{-2}}$ of the projections $P_0$ in eigenstate filtering.
As a result, this step requires at-most
\begin{eqnarray}
    && \order{M \Delta^{-2(1+\nu_l)} \log^{1+\nu_l}(1/\varepsilon_\mr{EF})} \nonumber \\
    && \quad = \order{\kappa^{2(1+\nu_l)} (\log \kappa)^2 (\log \log \kappa)^{1+\nu_l}} 
\end{eqnarray}
queries to the unitary gates $O_A$ and $U_b$.
In order to organize the block-encoding of $\tilde{A}(s)$ for every $s=0,1/M,2/M,\hdots,1$, we also need $M+1 \in \order{(\log \kappa)^2}$ different angles, which are analytically determined by $\arctan \sqrt{f_s/(1-f_s)}$.

After the reproduction of Zeno effect by eigenstate filtering, the algorithm applies the projection $P_0(\tilde{A})$ within the allowable error $\varepsilon$.
This step yields $\order{\Delta^{-2(1+\nu_l)} \log^{1+\nu_l}(1/\varepsilon)}= \order{\kappa^{2(1+\nu_l)}\log^{1+\nu_l}(1/\varepsilon)}$ query complexity.
Due to the $\order{1}$ overlap after Zeno effect, this step outputs the accurate solution of QSLP, $\mr{t}(\ket{x}, 0)$, with $\order{1}$ success probability probability.
In total, the QLSP algorithm exploiting eigenstate filtering with r-QET is executed by
\begin{equation}
    \tilde{\mcl{O}}\left(\kappa^{2(1+\nu_l)} \log^{1+\nu_l} (1/\varepsilon)\right)
\end{equation}
queries to the unitaries $O_A$ and $U_b$.
This scaling is worse than that of the original one using the standard optimal QET for eigenstate filtering \cite{Lin2020-filtering_qlsp}, whose scaling is $\tilde{\mcl{O}}(\kappa \log (1/\varepsilon))$.
However, the algorithm based on r-QET has a strong advantage in the parameter determination also in QLSP.
While the standard one requires $\tilde{O}(\kappa \log (1/\varepsilon))$ different parameters found by numerically-unstable computation, r-QET works only with $4l+3+M+1 \in \order{(\log \kappa)^2}$ different parameters, and all of them can be found analytically or in much numerically-cheaper ways.

\end{document}